\documentclass{iaus}
\usepackage{graphicx}
\usepackage{natbib}

\newcommand{\msun}{M$_{\odot}$}

\newcommand{\logg}{$\log{g}$}
\newcommand{\teff}{T$_{eff}$}

\newcommand{\lii}{Li~I}
\newcommand{\meth}{CH$_4$}
\newcommand{\wat}{H$_2$O}

\newcommand{\micron}{$\mu$m}

\newcommand\aj{AJ} 
\newcommand\araa{{ARA\&A}} 
\newcommand\apj{{ApJ}} 
\newcommand\apjl{{ApJL}} 
\newcommand\apjs{{ApJS}} 
\newcommand\aap{{A\&A}} 
\newcommand\nat{{Nature}} 
\newcommand\mnras{{MNRAS}} 

\title[Brown Dwarf Chronometers] 
{Brown Dwarfs as Galactic Chronometers}

\author[Adam J.\ Burgasser]   
{Adam J.\ Burgasser$^1$}

\affiliation{$^1$Massachusetts Institute of Technology, 
Kavli Institute for Astrophysics and Space Research, \\
Building 37, Room 664B, 77 Massachusetts Avenue, 
Cambridge, MA 02139, USA \\ email: {\tt ajb@mit.edu}}

\pubyear{2008}
\volume{258}  
\pagerange{XXXXXX}
\jname{Ages of Stars}
\editors{E.\ Mamajek, D. Soderblom \& J. Valenti, eds.}

\begin{document}

\maketitle

\begin{abstract}
Brown dwarfs are natural clocks, cooling and dimming over time due to insufficient core fusion.  
They are also numerous and present in nearly all Galactic environments, making them
potentially useful chronometers for a variety of Galactic studies.   For this potential to be realized, however,
precise and accurate ages for individual sources are required, a prospect made difficult by
the complex atmospheres and spectra of low-temperature brown dwarfs;
degeneracy between mass, age and luminosity; and the lack of useful age trends
in magnetic activity and rotation.  In this contribution, I review
five ways in which ages for brown dwarfs are uniquely determined, discuss their
applicability and limitations, and give current empirical precisions.  
\end{abstract}

\firstsection 
\section{Introduction}

Brown dwarfs are very low-mass stars
whose masses (M $\lesssim$ 0.075~{\msun}) are insufficient to sustain 
the core hydrogen fusion reactions that balance radiative energy losses \citep{1963ApJ...137.1121K,1963PThPh..30..460H}.  Supported from further gravitational contraction
by electron degeneracy pressure, evolved brown dwarfs continually cool
and dim over time as they radiate away their initial
contraction energy, ultimately
achieving photospheric conditions that can be similar to those of giant planets.  
The first examples of brown dwarfs were identified as recently as 
1995 \citep{1995Natur.378..463N,1995Natur.377..129R}.
Today, there are hundreds known in nearly
all Galactic environments, identified largely
in wide-field, red and near-infrared imaging surveys such as 2MASS, DENIS, SDSS and UKIDSS.
The known population of brown dwarfs encompasses
the late-type M ({\teff} $\approx$ 2500--3500~K), L ({\teff} $\approx$ 1400--2500~K) and T spectral classes ({\teff} $\approx$ 600--1400~K; e.g., \citealt{2004AJ....127.2948V}), 
while efforts are currently underway to find even colder
members of the putative Y dwarf class (see review by \citealt{2005ARA&A..43..195K}).

Because brown dwarfs cool over time, their spectral properties are inherently time-dependent,
making them potentially useful chronometers for Galactic studies (much like white dwarfs;
see contributions by M.\ Salaris, J.\ Kalirai, S.\ Catal{\'a}n and H.\ Richer).
However, the primary observables of a brown dwarf---temperature, luminosity and spectral type---depend on both mass and age (and weakly on metallicity).  This degeneracy complicates characterizations of individual sources and mixed
populations.  Unfortunately, traditional stellar age-dating methods do not appear to be applicable 
for brown dwarfs.  
Magnetic activity metrics, such as the frequency and strength
of optical or X-ray non-thermal emission, appear to be largely age-invariant
(e.g., \citealt{2006A&A...448..293S}) and quiescent emission drops off
precipitously in the early L dwarfs as cool photospheres are decoupled from
field lines
\footnote{Interestingly, radio emission does not drop off for cooler brown
dwarfs, and may even 
increase relative to bolometric flux, although there are currently few detections (e.g., \citealt{2006ApJ...648..629B}).}
(e.g., \citealt{2002ApJ...571..469M}; see contribution by A.\ West).  Long-term angular momentum loss in brown
dwarfs is far more muted than in stars, and there is no clear rotation-activity
relation for L dwarfs (e.g., \citealt{2008ApJ...684.1390R}).
Exploitation of the cooling properties of brown dwarfs is therefore a favorable approach
for determining their ages.

In this contribution, I review five methods currently employed to age-date brown dwarfs and summarize 
their applicability, inherent limitations and current (typical) precisions.
Table~1 provides a summary of the methods discussed in detail below.

\begin{table}
  \begin{center}
  \caption{Age-dating Methods for Brown Dwarfs.}
  \label{tab1}
 {\scriptsize
  \begin{tabular}{|l|l|l|l|c|l|}
 \hline 
{\bf Technique} & 
{\bf Applicable for} & 
{\bf Pros} & 
{\bf Cons} & 
{\bf Precision} & 
{\bf Examples} \\ 
 & & & & & 
{\bf in the} \\ 
 & & & & & 
{\bf Literature} \\ 
 \hline
Cluster & nearby clusters; &  precise ages based on  & generally restricted to & $\sim$10\% clusters; & 1,2,3,4 \\
members \&  & companions to & stellar/cluster work; & young clusters; wide & $\sim$50--100\% for & \\
companions  & age-dated stars & calibration for other & (resolvable) \& close & companions & \\
  & & techniques and & (RV) companions rare; & & \\
  & & evolutionary models &  must verify coevality/ & & \\
  & & &  association; atmospheres & & \\
  & & &  variable for $t \lesssim$ 10~Myr & & \\
 \hline
6708~{\AA} Li~I & $t$ $\sim$ 10--200~Myr; &  consistent predictions   & requires high sensitivity, & $\sim$10\% for  & 5,6,7 \\
absorption  & resolved binaries; & from different models; & high resolution spectra; & young clusters; & \\
  & individual field  & straightforward test; & low brightness region; & upper/lower & \\
  & sources with & largely insensitive to  & not useful for T dwarfs & limits for all & \\
  & {\teff} $>$ 1500~K & atmospheric properties & or for $t \lesssim$ 10~Myr; & others & \\
  & and $t \lesssim$ 2~Gyr  &  & relies on accurate & & \\
  & (limits only)  & & evolutionary models & & \\
 \hline
Mass & astrometric/RV & precise masses    & suitable systems are & $\sim$10--20\% & 8,9,10,11 \\
standards  & binaries & yield precise ages; & rare; long-term & & \\
  & &  weakly sensitive  & follow-up required; & & \\
  &  & to atmospheric & relies on accurate & & \\
  &  & properties & evolutionary models & & \\
 \hline
Surface & any source with & applicable to & low precision; other & $\sim$50--100\% & 12,13,14 \\
gravities  & a well-measured & individual sources;    & factors (e.g., metallicity) & & \\
 & spectrum & particularly useful & complicate analysis; & & \\
  & & for T dwarfs & relies on accurate & & \\
  &  & & evolutionary and & & \\
  &  & & atmospheric models & & \\
 \hline
Kinematics & well-defined & useful statistical test & very low precision; & $\sim$100--300\% & 15,16,17 \\
 & groups or & for various subclasses;   & large groups required & & \\
 & populations & insensitive to   & to beat statistical noise; & & \\
  &  & evolutionary or &  susceptible to & \\
  &  &  atmospheric models & selection biases & \\
 \hline
  \end{tabular}
  }
 \end{center}
\vspace{1mm}
 \scriptsize{
 {\it References:}
 (1) \citet{1998A&A...336..490B}; (2) \citet{2003ApJ...593.1093L}; (3) \citet{2001ApJ...556..373G};
 (4) \citet{2007ApJ...658..617B}; (5) \citet{1998ApJ...499L.199S}; (6) \citet{2005MNRAS.358...13J}; 
 (7) \citet{2005ApJ...634..616L}; (8) \citet{2004ApJ...615..958Z};
 (9) \citet{2006Natur.440..311S}; (10) \citet{2008arXiv0807.0238L}; (11) \citet{2008arXiv0807.2450D};
 (12) \citet{2004ApJ...609..854M}; (13) \citet{2006ApJ...639.1095B}; (14) \citet{2007ApJ...656.1136S};
 (15) \citet{2007AJ....133.2258S}; (16) \citet{2007ApJ...666.1205Z}; 
 (17) \citet{2008arXiv0809.3008F}.
 }
\end{table}

\section{Cluster Members and Companions}

The most straightforward way to age-date an individual brown dwarf is to borrow 
from its environment, a tactic that is suitable for members of coeval
clusters/associations and companions to age-dated stars.  
Brown dwarfs are well-sampled
down to and below the deuterium fusing mass limit (M $\lesssim$ 0.013~{\msun})
in the youngest nearby clusters ($t$ $\lesssim$ 5~Myr), as their
luminosities are greater at early ages. Brown dwarfs have
also been identified in somewhat older (10--50~Myr)``loose associations'' in the vicinity
of the Sun ($d$ $\lesssim$ 50--100~pc; e.g., \citealt{2008arXiv0808.3153K}).  
For older and more distant clusters ($d$ $\gtrsim$ 1~kpc),
decreasing surface temperatures and compact radii exacerbate the sensitivity
issues that plague low-mass stellar studies (see contribution by G.\ Piotto).  There are as yet
no known brown dwarfs in globular clusters, despite detection of the end 
of the stellar main sequence in systems such as NGC~6397 \citep{2008AJ....135.2141R}.

For brown dwarfs in young clusters, 
numerous studies have examined age-related trends in 
colors (e.g., \citealt{2008MNRAS.385.1771J}), 
spectral characteristics (e.g., \citealt{2007ApJ...657..511A}), 
accretion timescales (e.g., \citealt{2005ApJ...626..498M}) and
circum(sub)stellar disk evolution (e.g., \citealt{2007ApJ...660.1517S}).  These
have been coarsely quantified, and appear to be most useful at very 
young ages ($t \lesssim$ 10~Myr).
Surface properties and luminosities are highly variable at these ages
due to sensitivity to formation conditions
(e.g., \citealt{2002A&A...382..563B}), ongoing accretion,
complex magnetic effects (e.g., \citealt{2007ApJ...671L.149R})
and age spreads within a cluster (see contribution by R.\ Jeffries).  
Hence, while brown dwarfs in clusters with ages 
spanning $\sim$1--650~Myr are now well-documented, with age uncertainties
as good as 10\% (for the LDB technique; see $\S$~3), useful predictive trends are probably 
limited to ages of $\gtrsim$10~Myr.

Known brown dwarf companions to main sequence stars now number a few dozen, 
spanning a wide range of separation, age and composition.  Many of these systems are widely-separated so that their brown dwarf
companions can be directly studied.
Age uncertainties for companion brown dwarfs depend on stellar dating methods which are generally
more uncertain (50--100\%; e.g., \citealt{2008arXiv0807.0238L}) than cluster ages.  Searches for substellar companions to more precisely age-dated
white dwarfs (e.g. \citealt{2008MNRAS.388..838D,2008ApJ...681.1470F}) and subgiants \citep{2006MNRAS.368.1281P} have so far met with limited success.   
Nevertheless, brown dwarf companions to age-dated stars serve as important benchmarks for calibrating
other age-dating methods at late ages ($\gtrsim$500~Myr) and are fundamental
for testing evolutionary models (see contribution by T.\ Dupuy).

\section{6708~{\AA} {\lii} Absorption}

Lithium is fused at a lower temperature 
than hydrogen (2.5$\times$10$^6$~K), resulting in a somewhat lower fusing mass limit
(0.065~{\msun}; \citealt{1997ApJ...482..442B}).   
Because the interiors of low mass stars and brown dwarfs are fully convective at early ages, an
object with a mass above this limit will fully deplete its initial reservoir of lithium within $\sim$200~Myr.
Hence, any system older than this
which exhibits {\lii} absorption has a mass less than 0.065~{\msun}
and is therefore a brown dwarf (e.g., \citealt{1992ApJ...389L..83R}).  With a mass limit, one can use evolutionary models to determine an age limit.

In the age range $\sim$10--200~Myr, the degree of lithium depletion in low-mass stars
and brown dwarfs
is itself mass-dependent, occurring earlier in more massive stars which first achieve 
the necessary core temperatures.  Hence, over this range the age of an individual source can be
precisely constrained if its mass is known.  A more practical approach, however, is to
ascertain the age of a group of coeval low-mass objects based on which sources do or do not
exhibit {\lii} absorption; this is the lithium depletion boundary (LDB) technique.  Different
evolutionary models yield remarkably similar predictions for the location of the LDB over a broad
range of ages \citep{2004ApJ...604..272B}, and the boundary itself is readily identifiable in color-magnitude
diagrams.  As such, this technique has been used to age-date several nearby young clusters and 
associations (e.g., \citealt{1998ApJ...499L.199S,1999ApJ...522L..53B,2005MNRAS.358...13J,2008arXiv0808.3584M}).  LDB studies have also provided independent
confirmation of other cluster-dating methods such as isochrone fitting \citep{2005MNRAS.358...13J}.
A variant of the LDB technique for coeval binary systems has been proposed by \citet{2005ApJ...634..616L}, in which a system that exhibits {\lii} absorption in the secondary but not
in the primary can have both lower and upper age limits assigned to it (note that the presence/absence of {\lii} in both components simply sets a single upper/lower age limit).  This technique requires resolved optical spectroscopy of both components and can be pursued only for
a few (rare) wide low-mass pairs (e.g., Burgasser et al., in prep.) 
or using high spatial-resolution spectroscopy (e.g. \citealt{2006A&A...456..253M}). 
No single brown dwarf pair straddling the LDB has yet been identified.

Despite its utility, the detection of LI~I absorption in brown dwarf spectra has
limitations.  The 6708~{\AA} line lies in an relatively faint spectral region for cool L-type dwarfs, so sensitive spectral observations on large telescopes are typically required to 
detect (or convincingly rule out) this feature.  For optically-brighter
M-type brown dwarfs,
high-resolution observations are generally required to distinguish {\lii} absorption from overlapping 
molecular absorption features.  Young brown dwarfs ($t \lesssim$ 50~Myr)
with low surface gravities show weakened alkali line absorption (see $\S$~5), 
including {\lii}, making it again necessary to obtain sensitive, high-resolution
observations \citep{2008arXiv0808.3153K}.  For brown dwarfs cooler than $\sim$1500~K
(i.e., the T dwarfs), lithium is chemically
depleted in the photosphere through its conversion to LiCl, LiF or LiOH \citep{1999ApJ...519..793L}.  As such, practical age constraints using {\lii}
can only be made for systems younger than $\sim$2~Gyr.

\section{Mass standards}

One way of breaking the mass/age/luminosity degeneracy for an individual brown dwarf is to explicitly
measure its mass.  This is feasible for sufficiently tight brown dwarf binaries for which
radial velocity (RV) and/or astrometric orbits can be measured.  
Of the $\sim$100 very low mass (M$_1$,M$_2$ $\leq$ 0.1~{\msun}) binary systems
now known, only a handful have sufficiently short periods that
large portions of their RV orbits (e.g., \citealt{2007ApJ...666L.113J,2008ApJ...678L.125B}), astrometric orbits (e.g., \citealt{2001ApJ...560..390L,2004A&A...423..341B, 2008arXiv0807.0238L,2008arXiv0807.2450D}), or both \citep{2004ApJ...615..958Z,2006Natur.440..311S} have been measured.  With measurable total system masses or mass functions, individual masses can be
estimated from relative photometry or directly
determined from recoil motion in both components (e.g., \citealt{2006Natur.440..311S}).  The individual masses and component
luminosities can then be compared to evolutionary models to determine ages.

\citet{2008arXiv0807.0238L} have suggested that such ``mass standards'' provide more precise constraints on the 
physical properties (including ages) of brown dwarfs as compared to ``age standards'', namely companions to 
main sequence stars.  Orbital masses can currently be constrained to 
roughly 5--10\% precision, translating into 10--20\% uncertainties in ages based
on evolutionary tracks (versus 50--100\% for main-sequence stars).  
More importantly, brown dwarf binaries with
mass measurements and independent
age determinations---i.e., companions to age-dated stars and cluster members---can provide
specific tests of the evolutionary models themselves.  Further details are provided
in the contribution by T.\ Dupuy.  

\section{Surface gravity}

Only 10--20\% of brown dwarfs are found to be multiple (e.g., \citealt{2006ApJS..166..585B}) and few of these are suitable for orbital mass measurements.  A proxy for mass  
is surface gravity, which can be determined directly from a brown dwarf's spectrum.  
For {\teff} $\lesssim$ 2500~K and ages greater than $\sim$50~Myr, evolutionary
models predict that brown dwarf surface gravities ($g \propto M/R^2$) are roughly proportional to mass due to
near-constant radii (roughly equal to Jupiter's radius).
Surface gravity is also proportional to photospheric pressure ($P_{ph} \propto g/{\kappa}_R$, where $\kappa_R$ is the Rosseland mean opacity),  which in turn influences the chemistry, line broadening and (in some cases) opacities of absorbing species in the photosphere.
Hence, ``gravity-sensitive'' features in a brown dwarf's spectrum can be used to
infer its mass and, through evolutionary models, its age.

Examples of gravity-sensitive features include the optical and near-infrared VO bands and alkali lines in late-type M and L dwarfs, all of which evolve considerably between field dwarfs ({\logg} $\approx$ 5 cgs),
young cluster dwarfs ({\logg} $\approx$ 3--4) and giant stars ({\logg} $\approx$ 0; e.g., \citealt{1999ApJ...525..466L}).
Enhanced VO absorption and weakened alkali line absorption is a characteristic trait
of young brown dwarf spectra
(e.g., \citealt{2003ApJ...593.1074G,2007ApJ...657..511A}). 
Alkali features in particular are useful for cooler brown dwarfs 
as VO condenses out of the photosphere.
Quantitative analyses of these features have been used to distinguish ``young'' ($\gtrsim$10~Myr) from ``very young'' sources thus far ($\lesssim$5~Myr; e.g., \citealt{2007AJ....133..439C}).
More robust metrics await larger and more fully-characterized samples.

Another important surface gravity diagnostic is collision-induced H$_2$ 
absorption, a smooth opacity source spanning a
broad swath of the infrared (e.g., \citealt{1969ApJ...156..989L,1997A&A...324..185B}). 
H$_2$ absorption is weakened in the low-pressure atmospheres of
young cluster brown dwarfs, resulting in reddened near-infrared spectral energy
distributions and colors; in particular, a characteristic, triangular-shaped H-band (1.7~{\micron})
flux peak (e.g., \citealt{2001MNRAS.326..695L,2006ApJ...639.1120K}).
The proximity of many young and reddened brown dwarfs ($<$ 100~pc) rules out ISM absorption as a primary source for this reddening.   \citet{2008MNRAS.385.1771J} have exploited this trend by using a 
proper-motion selected sample of nearby young cluster candidate members to infer an 
age/color/luminosity relation for brown dwarfs younger than 0.7~Gyr, with a stated accuracy
of $\pm$0.2 dex in log(age), or about 60\% fractional uncertainty.  
Kinematically older low-mass stars and brown dwarfs in the Galactic disk 
(e.g., \citealt{2008arXiv0809.3008F}) and halo populations (e.g., \citealt{2003ApJ...592.1186B})
exhibit unusually blue near-infrared colors due to enhanced H$_2$ absorption.
However, differences in metallicities and condensate cloud properties can 
muddle surface gravity determinations in these sources by modulating the
photospheric pressure through opacity effects (changing $\kappa_R$; e.g.,
\citealt{2000ApJ...535..965L,2008ApJ...686..528L}).

The use of H$_2$ absorption as a surface gravity indicator is particularly useful
for T dwarfs, as H$_2$ dominates the $K$-band (2.1~{\micron}) opacity
and significantly influences near-infrared colors (e.g., \citealt{2002ApJ...564..421B,2004AJ....127.3553K}).
Several groups now employ this feature to estimate the atmospheric properties of individual T dwarfs
(e.g.,  \citealt{2006ApJ...639.1095B,2007MNRAS.381.1400W,2008MNRAS.tmp.1183B}), typically through the use of 
spectral indices that separately sample surface gravity (e.g., the $K$-band) and temperature variations (e.g., {\wat} or {\meth} bands).  These indices are compared
to atmospheric models calibrated by one or more benchmarks (e.g., a companion to a precisely age-dated star), and evolutionary models are used to determine individual masses and ages.
Typical uncertainties of {\logg} $\sim$0.3~dex translate into 50-100\% uncertainties in age, comparable to uncertainties for main
sequence stars.
Again, variations in metallicity can mimic variations in surface gravity, although a third
diagnostic such as luminosity can break this degeneracy (e.g., \citealt{2007ApJ...658..617B}). 
As atmosphere models improve in fidelity, parameters are increasingly inferred
from direct fits to spectral data, with comparable uncertainties
(e.g., \citealt{2007ApJ...656.1136S,2008ApJ...678.1372C}).

\section{Kinematics}

When a sufficiently large enough sample of brown dwarfs is assembled, one can
apply standard kinematic analyses, building from the assumption 
that gravitational perturbations lead to increased velocity
dispersions with time (e.g., \citealt{1953ApJ...118..106S}; see
contribution by B.\ Nordstr\"om). Velocity dispersions are typically characterized
by a time-dependent power-law form, i.e., $\sigma \propto (1+t/\tau)^{\alpha}$ (e.g., \citealt{1977A&A....60..263W,2002MNRAS.337..731H}). Other statistics, such as Galactic scale height, can also be tied to age through kinematic simulations (e.g., \citealt{2008AJ....135..785W}) to calibrate
secondary age diagnostics such as magnetic activity (see contribution by A.\ West).

Samples of very low mass stars and 
brown dwarfs have only recently become large enough that kinematic studies are feasible.
The largest samples (over 800 sources) have been based on proper motion measurements (e.g., \citealt{2007AJ....133.2258S,2008MNRAS.390.1517C,2008arXiv0809.3008F}).  For field dwarfs,
dispersion in tangential velocities for both magnitude- and volume-limited samples indicate a mean age
in the range 2--8~Gyr, largely invariant with spectral type.  This is consistent with the ages of
more massive field stars and population synthesis models
(e.g., \citealt{2004ApJS..155..191B}).  However, when field samples are broken down by color \citep{2008arXiv0809.3008F} or presence of magnetic activity \citep{2007AJ....133.2258S}, distinct age groupings are inferred, indicating that both very old (i.e., thick disk or halo) and very young (i.e., thin disk or young association) brown dwarf populations coexist in the immediate vicinity of the Sun.  Indeed, one of the major results from these studies is the identification of widely-dispersed brown dwarf members of nearby, young moving groups such as the Hyades (e.g., \citealt{2007MNRAS.378L..24B}).

With only two dimensions of motion measured, proper motion samples may produce biased dispersion measurements depending on the area of sky covered by a sample.  Full 3D velocities require 
RV measurements which are more expensive and have thus far been obtained only for a small fraction of the known brown dwarf population
(e.g., \citealt{2006AJ....132..663B,2007ApJ...666.1198B}).
A seminal study by \citet{2007ApJ...666.1205Z} of 
21 nearby, late-type dwarfs with parallax, proper motion and RV measurements found 
considerably smaller 3D velocity dispersions for L and T dwarfs than GKM stars, suggesting that local brown dwarfs
are young ($t \sim$ 0.5--4~Gyr).  The discrepancy between this result
and the proper motion studies may be attributable to small number statistics and/or contamination
by young moving groups; $\sim$40\% of the brown dwarfs in the \citet{2007ApJ...666.1205Z} sample appear
to be associated with the Hyades.  Resolving this discrepancy requires
larger RV samples, which has the side benefit of potentially uncovering RV variables
that can be used as mass standards (e.g., \citealt{2006AJ....132..663B}).

\section{Improvements and Future Work}

With several methods for age-dating brown dwarfs over a broad range of phase space now
available, 
opportunities to use these objects as chronometers for various 
Galactic studies look to be increasingly promising; e.g., age-dating
planetary systems, examining cluster age spreads, 
testing Galactic disk dynamical heating models, and direct measures of the 
substellar mass function and birthrate in the field and other populations.  However, there are areas where improvements
in uncertainties are needed and basic assumptions tested.  Surface gravity determinations 
in particular require better constraints, since these 
enable age-dating of individual sources.  In the short term, improvements
in spectral models can help in this endeavor; however, a sufficiently sampled grid of 
benchmark sources  
may obviate the need for models entirely.
Benchmarks should increasingly arise from mass standards, for which
age constraints are more precise; these additionally provide
necessary tests of evolutionary models upon which most of the age-dating
techniques hinge.    Improved angular resolution and sensitivity
with $JWST$ and the next generation of large ($>25$m) telescopes
will increase resolved binary sample sizes by expanding the volume in which they
can be found and monitored.  These facilities will also aid searches for 
substellar cluster members in old open field and globular clusters and, perhaps more
importantly, mass standards in these clusters to facilitate more stringent tests of evolutionary models
(T.\ Dupuy, priv.\ comm.).
Finally, a larger, more complete sample of brown dwarfs with precise RV measurements
will both improve our statistical constraints on the age of the local brown dwarf
population (and subpopulations) while additionally aiding in the search for mass standards.

\bigskip

\noindent The author thanks T.\ Dupuy, S.\ Leggett, M.\ Liu, \& A.\ West for helpful comments.







\end{document}